# Lithium adsorption properties of monolayer $B_5Se$

Amretashis Sengupta[1*]

**\*Corresponding author e-mail:** amretashis@nbu.ac.in

University of North Bengal, Raja Rammohunpur, Dist. Darjeeling, West Bengal -734 013, India

**Abstract:**

*In this work we investigate the Li adsorption properties of a hybrid 2D material, namely monolayer $B_5Se$ with first principles calculations. 2 dimensional $B_5Se$ was found to have a distorted hexagonal structure with five B atoms and one Se atom at the vertices of each hexagon. The density functional theory (DFT) calculations are performed with the generalized gradient approximation (GGA) and Perdew-Burke-Ernzerhoff (PBE) exchange correlation functional. The results were inclusive of van der Waals corrections with Grimme's DFT-D2 scheme. Electrode performance metrics, as the most preferred adsorption sites and adsorption energies, open circuit anode potentials, charge density differences and specific capacities for varying adatom coverage were evaluated with the DFT calculations. The adatom diffusion barriers were evaluated with a nudged elastic band (NEB) method. The ab-initio calculations predict a maximum theoretical specific capacity of 1486.87 mAhg$^{-1}$ for Li adsorption on 2D $B_5Se$, which is over four times that of conventional Li ion battery anode materials. This coupled with an open circuit anode potential of 0.291-0.179V for different degrees of Li coverage, and a small Li diffusion barrier of 0.15eV, metallic nature of the sheet under pure and lithiated conditions and good charge density variations, make monolayer $B_5Se$ a potent anode material for Li-ion battery applications.*

**Keywords: Density functional theory, B5Se, 2D materials, Li-ion battery**

## I. Introduction

In recent years, layered two dimensional (2D) materials have amassed significant interest in the quest for improving electrode performance in Li-ion batteries (LIB). [1]-[4] Pure graphene suffers from the issue of Li adatom clustering, [5]-[8] and can at best be applicable as an additive to enhance electrode conductivity or for structural support. [1]-[8] In this regard, reduced graphene oxide (rGO) has shown much better performance as compared to pure graphene [6,8] where a high number of defect sites provide enhanced specific capacity around 1000 mAhg$^{-1}$. [2] Of late beyond graphene 2D materials such as MXenes [1,2,9], transition metal dichalcogenides (TMD) [10]-[13], Borophene [14,15], Phosphorene [16], Silicene, Germanene [17,18], MnO2 [19] , Mo2C [20] , VS2 [21], AlN [22], antimonene [23], Stanene [24] etc. have shown high specific capacities and low diffusion barriers, suitable for LIB applications. Black phosphorous is one such 2D material which has shown a very high specific capacity of 2596 mAhg$^{-1}$, but owing to high volume expansion in excess of 300% during Lithiation cycles, is prone to anode pulverization. [2] TMDs as $MoS_2$ have sufficient interlayer gaps, thus promoting easy intercalation and diffusion of Li ions (diffusion barrier of ~ 0.21eV), and capacity around 670-1000 mAhg$^{-1}$. [1,2,4,10]-[13] When combined with n-doped graphene, $WS_2$ also shows a very good capacity for Li-ion of 905 mAhg$^{-1}$ and excellent stability (no change in capacity even after 100 cycles). [1]-[4] 2D Boron or Borophene, as it is popularly known as, has a very high specific capacity of ~1300 mAhg$^{-1}$ and a very small diffusion barrier of about 0.03eV for Li ions, [1]-[4],[14,15] and thus acts as a point of interest pertaining to Boron based hybrid 2D materials for LIB applications. As the library of 2D materials is expanding with the synthesis/ isolation of new materials and structures, the investigation of various 2D materials for potential applications in the field of Li-ion battery technology, is still very relevant.

In this work we investigate the Li adsorption properties of a hybrid 2D material, namely 2 dimensional hexagonal $B_5Se$ sheets. With ab-initio calculations carried out with density functional theory (DFT), we evaluate the Li adatom adsorption properties of such a system, with a specific attention to the parameters important to novel anode materials. Our calculations involve the identification of the most preferred adsorption sites, evaluation of the theoretical binding energies and open circuit anode potential and the theoretical specific capacities of the material for varying degrees of lithiation. The diffusion barrier for Li atom on a monolayer $B_5Se$ sheet was also calculated with nudged elastic band (NEB) simulations. We also computed the charge density differences for varying degrees of adatom coverage and also investigated the variations in the density of states (DOS) for changes in conductive properties of the Li covered sheets.

## II. Methods

First-principles calculations were performed using Quantum ESPRESSO package implemented in Materials Square. [25,26] For our calculations we employed the generalized gradient approximation (GGA) with the Perdew –Burke-Ernzerhoff (PBE) exchange correlation functional. [27] Ultrasoft pseudopotentials for all the elements (B, Se and Li) from the standard solid-state pseudopotentials (SSSP) library were used in the calculations. [28] The wave function cut-off energy was set to 30Ryd and density cut-off energy was set to180Ry. [25] The supercell was sampled with a 3x3x1 Monkhorst-Pack k-point grid. [29] A Marzari-Vanderbilt cold smearing was used to determine the electronic occupations around the Fermi level. [30] The electron convergence threshold was taken to be $10^{-6}$ Ry, and the Davidson algorithm [31] was used for the diagonalization of the Kohn-Sham Hamiltonian for the SCF calculations, with mixing beta of 0.5. Van der Waals' corrections were included in the DFT calculations with the Grimme's DFT-D2 scheme. [32] For the structural relaxations, the Hellmann-Feynman forces were minimized below 0.01 ev/Å, employing the Broyden-Fletcher-Goldfarb-Shanno (BFGS) algorithm. [33]

The cohesive energy of the $B_5Se$ is evaluated as [34]-[37]

$$E_{Co} = (E_B - E_{Se} - E_{B5Se})  \quad (1).$$

In (1) $E_B$, $E_{Se}$ are the energies of the isolated Boron and Selenium atom and $E_{B5Se}$ is the energy of the compound.

The lithiation reaction at the anode can be expressed as

$$Li_x B_5 Se \leftrightarrow Li_x^+ B_5 Se + xe^- \quad (2).$$

The adsorption energy per atom, for the process of adsorption of $n$ number of Li adatoms can be calculated as [34]-[37]

$$E_{ads} = \frac{E_{Li_x B_5 Se} - E_{B_5 Se} - nE_{Li}}{n} \quad (3).$$

In (3), $E_{Li_x B_5 Se}$ stands for the total energy of the Li adsorbed sheet, $E_{B_5 Se}$ for the bare (without adatom) sheet and $E_{Li}$ is the energy of an isolated Li atom in the BCC configuration. [34,35]

Assuming the change in the internal energy to be the equal to the change in the Gibbs' free energy of the system, the open circuit anode voltage or OCV ($V_A$) may be expressed as [34]-[37]

$$V_A = \frac{-\Delta G}{xzF} = \frac{-\Delta E_{int}}{xzF} \quad (4).$$

With $z=1$ for Li and $F$ being the Faraday constant. The specific capacity of the material can be expressed as [34]-[37],

$$C_{sp} = \frac{xzF}{MW} \quad (5)$$

In (5), $MW$ is the molecular weight of the lithiated sheet having the chemical formula $Li_x B_5 Se$. The diffusion barriers for different possible diffusion pathways for Li adatom on the surface of the $B_5Se$ sheet were calculated with the nudged elastic band (NEB) method. [38,39] For NEB calculations we employed the projected velocity verlet optimisation scheme for the images. [38,39]

## III. Results & Discussions

Our calculations show that upon structural optimization, with a variable cell relaxation, monolayer $B_5Se$ takes a slightly distorted 2D hexagonal form, with the five B atoms and one Se atom at the vertices of each hexagon, as shown in Fig. 1(a). The 2D $B_5Se$ primitive cell is also shown in Fig. 1(b), having lattice parameters a=5.26Å and b=5.35Å. The hybrid 2D sheet was found to be of metallic nature as observed from the density of states (DOS) plot, presented in Fig. 1(c).

Thereafter we look to identify the most preferred Li adsorption sites on 2D $B_5Se$. For this purpose four different sites labelled, site-I, site-II, site-III and site-IV were examined, as shown in Fig. 2(a). Site-I sits above the Se atom, site-II is directly above the centre of the hexagon, and site –III and IV lie above two of the non-equivalent B atoms (due to the distortion of the honeycomb nature) at the edges of the hexagon.

Upon structural relaxations, it was found that site-II was the most favourable and stable Li adsorption site, both in terms of the total energy and the adatom binding energy. The relative total energy of the other sites with reference to site-II is shown in Fig. 2(b), while the adatom binding energies for the different sites are shown in Fig. 2(c). As per the sign convention used for the calculation, the most negative binding energy represents the most strongly bound system.

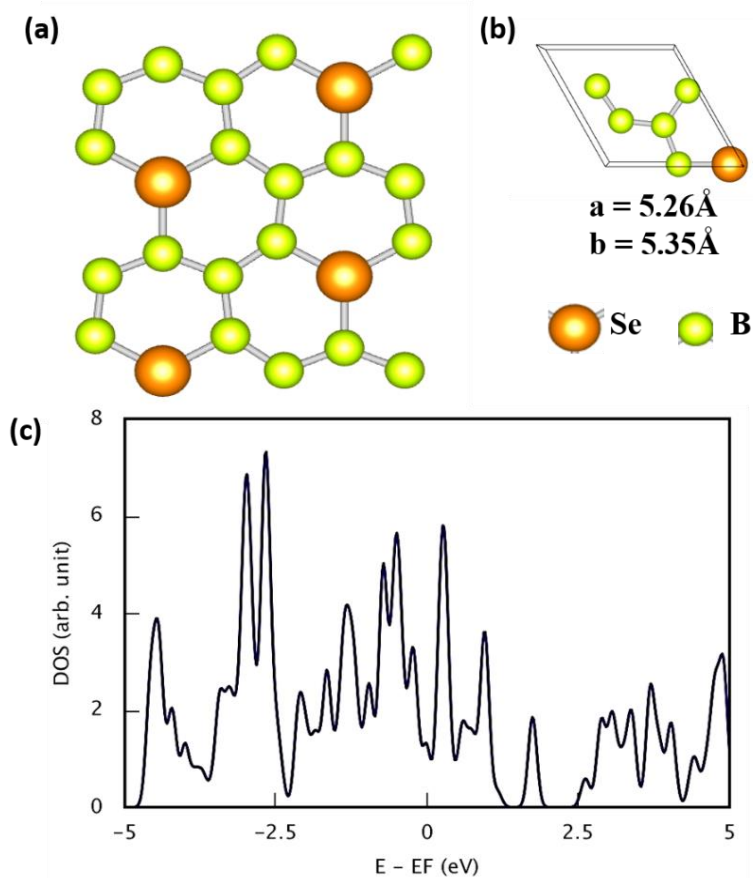

**Fig. 1:** (a) the optimized $B_5Se$ supercell used in the calculations (b) the hexagonal primitive cell for 2D $B_5Se$ and (c) the density of states (DOS) for the optimized structure, showing a metallic nature of the hybrid sheet.

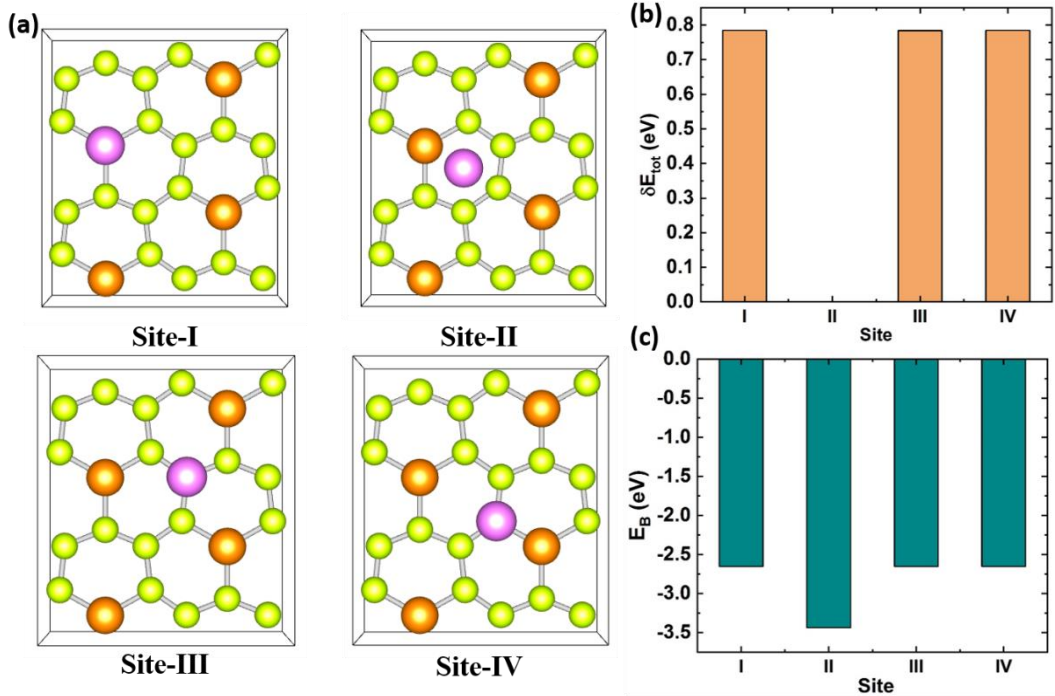

**Fig. 2: (a)** The different Li adatom adsorption sites **(b)** the relative total energies and **(c)** the binding energies of the different adsorption sites.

Having identified the most preferred Li adsorption site, we now proceeded towards calculations on various degrees of Li adatom coverage. For this study, the Li atoms are gradually filled on the available adsorption sites, and each structure is relaxed with the same parameters as mentioned in the methodology section. Through calculation of the total energies of the relaxed structures in each case, and thereafter using the formulae mentioned in the previous section, we calculate the adsorption (binding) energies per atom, the open circuit anode voltages and the specific capacities for different Li adatom coverage.

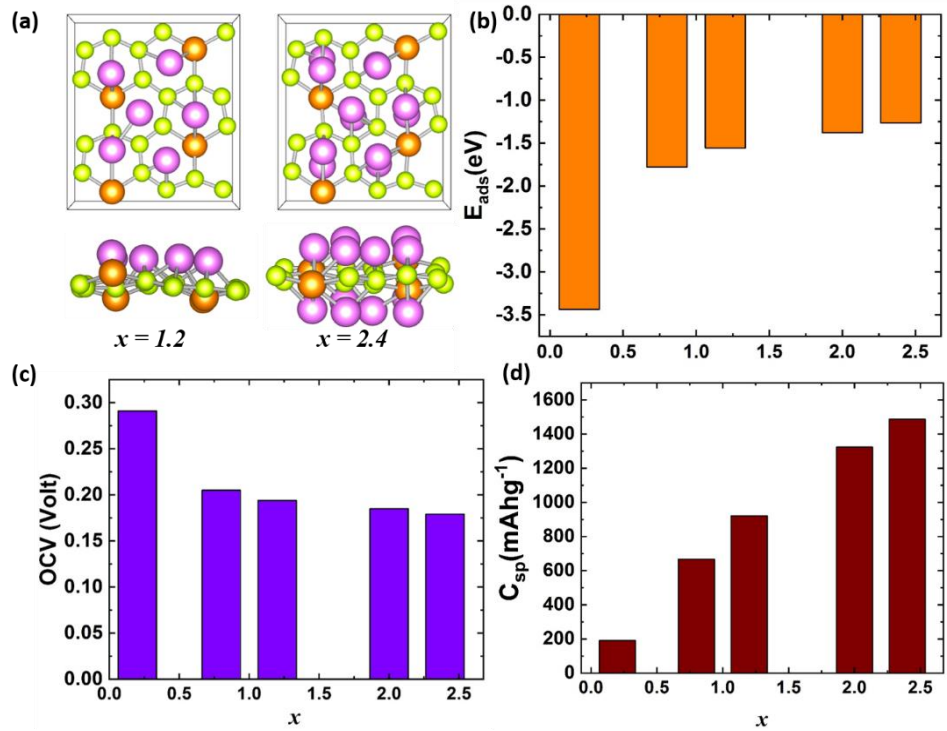

**Fig. 3: (a)** The top and side views of the lithiated B$_5$Se supercell, with all the preferred sites occupied on one ($x=1.2$) and both the sides ($x=2.4$) **(b)** the adsorption energy per atom **(c)** the open circuit anode voltage (OCV) and **(d)** the specific capacities for different Li adatom coverage.

On each side of the supercell, there exist six most preferred adsorption sites, making for a chemical formula of Li$_{1.2}$B$_5$Se for adatom coverage on one side of the sheet and Li$_{2.4}$B$_5$Se for the fully lithiated case. The optimized structures for these conditions ($x=1.2$ & $x=2.4$) are shown in Fig. 3(a). The relaxed position of the adsorbed Li atoms in such configurations are between 2.36-2.54Å above/ below the centre of the hexagons, with reference to plane of the 2D B$_5$Se sheet. Our calculations show that the adsorption energy per atom goes on increasing (considering the negative sign) from -3.44eV to -1.26eV as the Li adatom coverage increases, from $x=0.2$ to $x=2.4$. Comparatively Borophene sheets of similar Li coverage (x=0.25) have reported theoretical binding (adsorption) energies of -2.027eV.[15] These values indicates a stronger adsorption of Li adatom on B$_5$Se, as compared to Borophene and therefore less chances of Li clustering at similar levels of adatom coverage.

In terms of the open circuit anode voltage, shown in Fig. 3(c) a voltage of 0.29 to 0.18Volt was observed for different degree of Li adatom coverage. Such a voltage is lower than the average operating voltage of 0.44Volt for Borophene [15] and therefore makes B$_5$Se as a suitable anode material candidate for LIB. In terms of specific capacity, a high specific capacity going up to 1486.88 mAhg$^{-1}$ is displayed by B$_5$Se. This value is higher than that reported for Borophene, which is around 1239 mAhg$^{-1}$. Even when only one side of the B$_5$Se sheet is lithiated (i.e. $x=1.2$), the $C_{sp}$ has a value of 920.68 mAhg$^{-1}$, which is several times higher than that of conventional anode materials. As compared to other 2D materials such as MoS$_2$, WS$_2$, Mo$_2$C, VS$_2$, Stanene etc., [10]-[13],[20,21,24] the maximum theoretical specific capacity of 1486.88 mAhg$^{-1}$ calculated for B$_5$Se is found to be significantly higher.

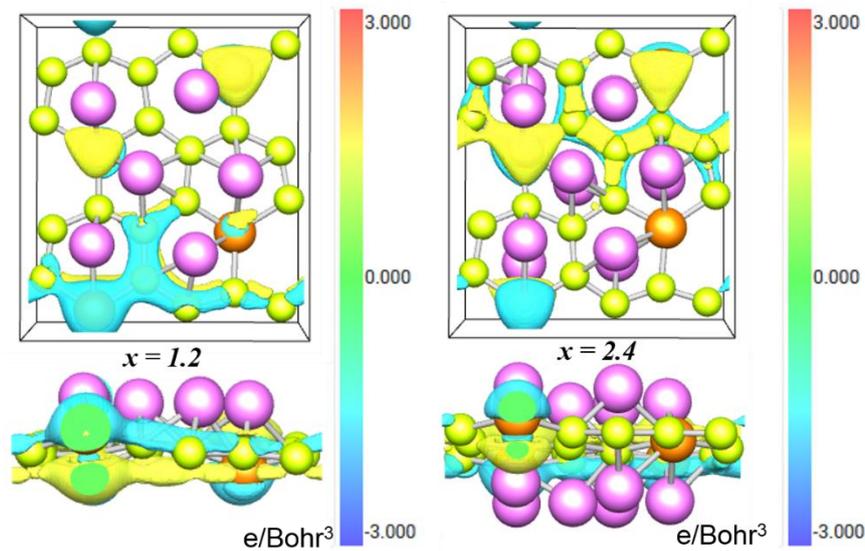

**Fig. 4:** Charge density difference plots for lithiated B$_5$Se for lithiation on one side (*x=1.2*) and both sides (*x=2.4*) of the sheet respectively.

The charge density difference (CDD) plots for the lithiated sheets for *x=1.2* and *x=2.4*, are shown in Fig. 4. The CDD is calculated as the difference between the charge densities of the bare B$_5$Se sheet and those of the lithiated ones. The data was plotted using an isovalue of 0.05e/Bohr$^3$. A Blue-Green-Red (BGR) colour scheme was employed to visualize the positive and the negative CDD regions. It is observable from the plots that for x=1.2, in the region between the adsorbed Li atoms and the B atoms, a predominantly negative CDD is visible, as indicated by the blue coloured isosurface. However, the Se atoms act as centres of mostly positive CDD, as in this case seen by the yellow blobs around most of the Se atoms. On the side of the B$_5$Se sheet without adatoms a region of positive CDD is also seen. When the sheet is fully lithiated (i.e. x=2.4), the presence of negative CDD region spreads to both sides of the sheet in regions between the adsorbed Li and B$_5$Se. The distribution of positive and negative CDD regions is however slightly sporadic considering the minor deformations of the 2D sheet with increasing Li coverage. A positive value of the CDD implies an accumulation of valence electrons, while a negative value of CDD implies the depletion of the same. From the CDD studies, we can understand that the Se atoms act as electron acceptors for the possible charge transfer, while depletion occurs around the B atoms, upon lithiation of the B$_5$Se sheet.

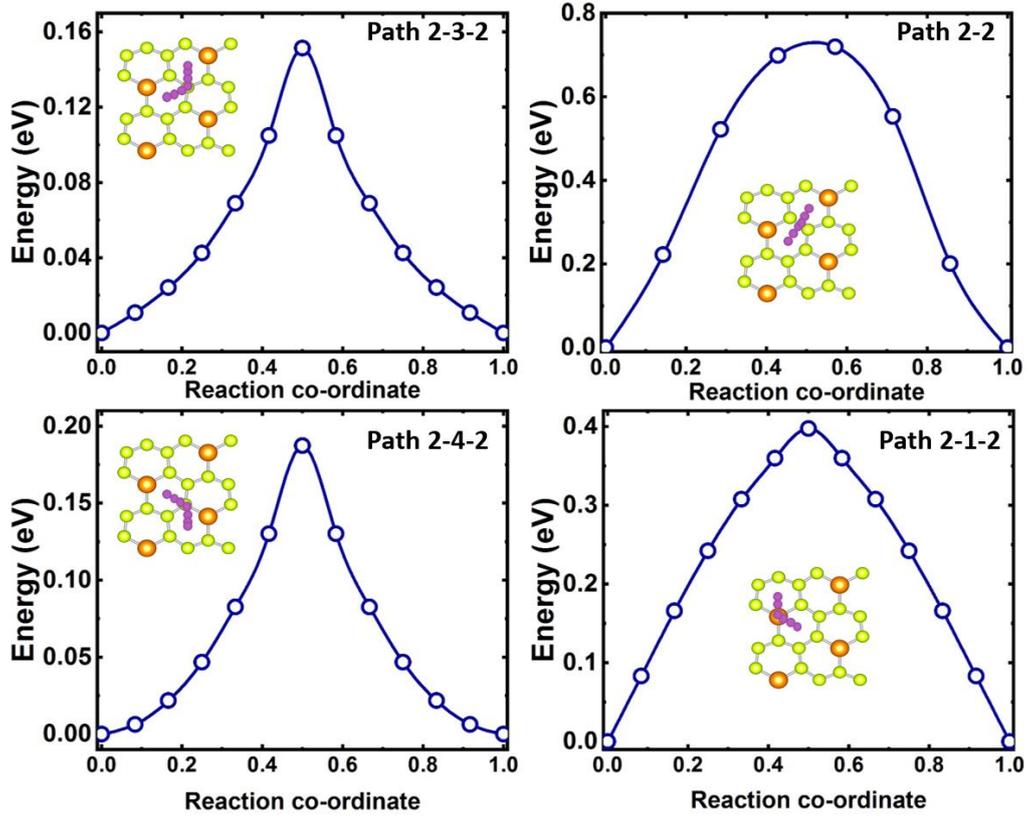

**Fig. 5:** Relative energy variations for different possible diffusion pathways (shown in insets), obtained with nudged elastic band (NEB) calculations.

In Fig. 5, we show the relative energy variations as obtained from NEB calculations. We consider the various possible pathways for the adsorbed Li atoms to move on the surface of the $B_5Se$ sheet from one most preferred site (site-II) to another equivalent site. Clockwise from the top left tile in Fig. 5, let's name these paths are path 2-3-2 (site II-site III-site II), path 2-2 (site II –site II), path 2-1-2 (site II-site I-site II), and path 2-4-2 (site II-site IV-site II) respectively. Our NEB calculations show that path 2-3-2, shows the smallest diffusion barrier ($\Delta$) of 0.151eV. This is followed closely by the path 2-4-2, which has a barrier height of 0.187eV. Path 2-2 and path 2-1-2 show larger diffusion barrier heights of 0.728eV and 0.397eV respectively. Thus the most preferred pathway for the Li adatom diffusion on 2D $B_5Se$ happens to be along path 2-3-2, in which the adatom moves from one site-II to another via site-III. This value of $\Delta$ =0.151eV, is quite smaller than that for a number of other 2D materials such as $MoS_2$, $VS_2$, Antemonene, AlN, Stanene etc. [10]-[13],[20,21,24]

An important criteria for electrode materials, remains its conductivity under pure (unlithiated) and the lithiated condition. From the density of states (DOS) plots of the $B_5Se$ sheet under different Li coverage, we see that the 2D $B_5Se$ retains its metallic nature as shown (Fig. 1c) in pure condition, upon lithiation as well, as evident from (Fig. 6) the non-zero DOS at the Fermi level. Thus, in this regard as well, 2D $B_5Se$ meets the criteria of a possible anode material for Li ion battery applications.

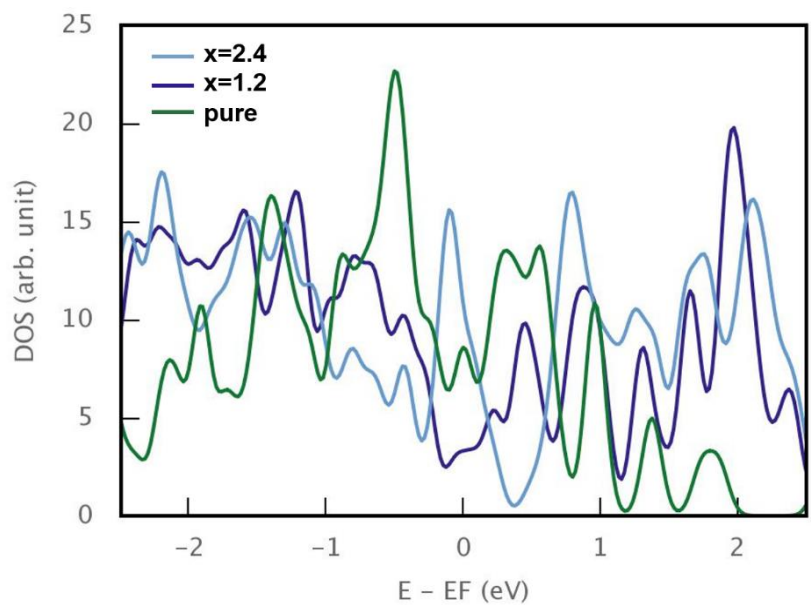

**Fig. 6:** The density of states (DOS) of the 2D $B_5Se$ for pure and lithiated conditions (*x=1.2* and *x=2.4*). The Fermi energy is set at zero, showing the metallic nature of the material.

## IV. Conclusion

In this work with density functional theory (DFT) calculations we investigate the hybrid 2D material $B_5Se$ for its possible applications as an anode material in lithium ion batteries. With DFT calculations using GGA-PBE exchange and correlation functional, and inclusive of Grimme's DFT-D2 corrections, structural relaxations and total energy calculations were carried out. From these most preferred Li adsorption site was identified, and electrode application metrics such as average adsorption energy, open circuit anode voltage (OCV), specific capacitance etc. were found out for different amounts of Li coverage. The DFT calculations showed a high Li adsorption energy of -3.44eV on $B_5Se$, with OCV values between 0.29-0.18V for different amounts of adatom coverage, and very high maximum theoretical specific capacity up to 1486.88 mAhg$^{-1}$. Nudged elastic band (NEB) calculations show that the minimum diffusion barrier for Li on 2D $B_5Se$ is only around 0.15eV. The results of our calculations show 2D $B_5Se$ to be a promising layered material for anode applications in Li-ion batteries.

**Acknowledgement**

The author thanks the Science and Engineering Research Board (SERB), Government of India for providing financial support under the SERB Research Scientist scheme, Grant No. SB/SRS/2019-20/03/ES.9